\documentclass[a4paper,11pt]{article}
\pdfoutput=1 

\usepackage{jheppub} 

\usepackage[T1]{fontenc} 
\usepackage{comment}
\usepackage{soul}

\usepackage[textsize=footnotesize,textwidth=2.cm]{todonotes}

\def\p{\partial}

\newcommand{\ba}{\begin{array}}
\newcommand{\ea}{\end{array}}
\newcommand{\bi}{\begin{itemize}}
\newcommand{\ei}{\end{itemize}}
\newcommand{\bea}{\begin{eqnarray}}
\newcommand{\eea}{\end{eqnarray}}
\newcommand{\be}{\begin{equation}}
\newcommand{\ee}{\end{equation}}

\title{\boldmath Structure Constants from Modularity in Warped CFT}

\author[\dagger]{Wei Song}
\author{and}
\author[*]{Jianfei Xu}
\date{\today}

\affiliation[\dagger]{Yau Mathematical Sciences Center,Tsinghua University, Beijing, 100084, China}
\affiliation[*]{Shing-Tung Yau Center and School of Mathematics, Southeast University, Nanjing, 211189, China}

\emailAdd{wsong2014@mail.tsinghua.edu.cn}
\emailAdd{jfxu@seu.edu.cn}

\abstract{
We derive a universal formula for the asymptotic growth of the mean value of three-point coefficient for Warped Conformal Field Theories (WCFTs), and provide a holographic calculation in BTZ black holes.
WCFTs are two dimensional quantum field theories featuring a chiral Virasoro and U(1) Kac-Moody algebra, and are conjectured to be holographically dual to quantum gravity on asymptotically AdS$_3$ spacetime with Comp$\grave{\mathrm{e}}$re-Song-Strominger boundary conditions.
The WCFT calculation amounts to the calculation of one-point functions on torus, whose high temperature limit can be approximated by using modular covariance of WCFT, similar to the derivation of Cardy formula. The bulk process is given by a tadpole diagram, with a massive spinning particle propagates from the infinity to the horizon, and splits into particle and antiparticle which annihilate after going around the horizon of BTZ black holes.
The agreement between the bulk and WCFT calculations indicates that the black hole geometries in asymptotically AdS$_3$ spacetimes can emerge upon coarse-graining over microstates in WCFTs,
similar to the results of Kraus and Maloney in the context of AdS/CFT~\cite{Kraus:2016nwo}.
}

\begin{document}
\maketitle
\flushbottom
\section{Introduction}

Field theories with $SL(2,R)\times U(1)$ global symmetry are of great interest from many perspectives. One motivation comes from holography for a large class of geometries with $SL(2, R)\times U(1)$ isometry, including the near horizon of extremal Kerr(NHEK)~\cite{Bardeen:1999px, Guica:2010ej} and  the warped AdS$_3$ (WAdS)~\cite{Anninos:2008fx} spacetime. Such geometries are not asymptotically locally AdS space, and thus the conjectured Kerr/CFT \cite{Guica:2008mu}, WAdS/CFT \cite{Anninos:2008fx}, WAdS/WCFT \cite{Detournay:2012pc}, AdS$_3$/WCFT \cite{Compere:2013bya} correspondence explores properties of holographic dualilty beyond the standard AdS/CFT correspondence.
Both the asymptotic analysis~\cite{Compere:2008cv,Compere:2009zj,Compere:2013bya,Compere:2014bia} in the bulk and the field theoretical analysis~\cite{Hofman:2011zj} indicates that the field theory with $SL(2,R)\times U(1)$ global symmetry have infinite dimensional local symmetries.
One minimal possibility is the warped conformal field theory  (WCFT) \cite{Detournay:2012pc} featuring Virasoro-Kac-Moody symmetries, and the other is to have both left and right Virasoro symmetries. Further discussions of the two possibilities can be found in~\cite{ElShowk:2011cm, Song:2011sr, Azeyanagi:2012zd,Compere:2014bia}, and in particular the recent developments on J${
\bar T}$ deformations of CFTs~\cite{Guica:2017lia, Bzowski:2018pcy, Chakraborty:2018vja, Apolo:2018qpq}.

In this paper, we focus on the WCFTs, which are two dimensional, nonrelativistic quantum field theory characterized by a Virasoro algebra and one U(1) Kac-Moody algebra~\cite{Hofman:2011zj, Detournay:2012pc}. Specific models of WCFT include chiral Liouville gravity~\cite{Compere:2013aya}, free Weyl fermions~\cite{Hofman:2014loa,Castro:2015uaa}, free scalars~\cite{Jensen:2017tnb}, and also the Sachdev-Ye-Kitaev models with complex fermions~\cite{Davison:2016ngz} as a symmetry-broken phase~\cite{Chaturvedi:2018uov}.
Data of WCFTs are the spectrum of operators and the three-point function coefficients or the structure constants. Using warped conformal symmetry and Operator Product Expansion (OPE), an arbitrary correlation function of a WCFT can be constructed in terms of these data.
Similar to the conformal bootstrap~\cite{Ferrara:1973yt, Polyakov:1974gs, Belavin:1984vu},  a warped conformal bootstrap program has been initiated in~\cite{Song:2017czq} using crossing symmetry.

In the holographic dual, the Virasoro-Kac-Moody algebra arises from the asymptotic symmetry analysis under Dirichlet-Neumann boundary conditions.
Examples include topologically massive gravity on warped AdS$_3$~\cite{Compere:2008cv,Compere:2009zj}, IIB supergravity on a class of solutions containing a AdS$_3$ factor \cite{Song:2011sr, Detournay:2012dz, Compere:2014bia}, and Einstein gravity on AdS$_3$ \cite{Compere:2013bya}. All these examples with a metric formulation feature 
 a negative Kac-Moody level as well as a purely imaginary vacuum charge \cite{Detournay:2012dz} \footnote{In purely field theoretical analysis for unitary WCFTs have positive central charge and Kac-Moody level.  A holographic dual to unitary WCFTs has also been proposed in~\cite{Hofman:2014loa}, where the bulk theory is a Chern-Simons theory called the lower spin gravity, but it does not have a metric formulation. In this paper, when we say ``holographic WCFTs'' we exclude this example.  }.
Evidence of the aforementioned AdS/WCFT and WAdS/WCFT correspondence accumulates and one finds that the duality is compatible and sometimes only compatible with the negative level.
In particular, WCFTs are covariant under modular transformations~\cite{Detournay:2012pc, Castro:2015csg}. As a result, a Cardy-like asymptotic formula of density of states~\cite{Detournay:2012pc} has been derived, which successfully matches the Bekenstein-Hawking entropy of black holes in the bulk.
Entanglement entropy for WCFTs was first calculated in \cite{Castro:2015csg}. To find the bulk dual, one needs to perform a more general Rindler transformation~\cite{Song:2016gtd} compatible to the non-vanishing vacuum charge~\cite{ Song:2016gtd,Azeyanagi:2018har, Chen:2019xpb}. The resulting geometric picture is different from the Ryu-Takayanagi proposal.
Besides a negative level $k$ and a pure imaginary vacuum charge \cite{Detournay:2012dz}, holographic WCFTs also feature vacuum block dominance \cite{Chen:2019xpb, Apolo:2018oqv} and maximal chaos \cite{Apolo:2018oqv} at a large central charge $c$.

In this paper, we study another universal property of holographic WCFTs, the asymptotic growth of three point coefficients, along the lines of~\cite{Kraus:2016nwo}. In~\cite{Kraus:2016nwo}, asymptotic formula of the average three-point coefficient in the heavy-heavy-light limit for scalar operators was derived in CFT$_2$. Using  the AdS$_3$/CFT$_2$ dictionary, the aforementioned quantity can be reproduced by a bulk tadpole diagram calculation on BTZ black hole background. In particular, the result can be approximated using the geodesic of massive point particles. The goal of this paper is to understand if such a picture is still valid in the proposed holographic duality between WCFT and Einstein gravity with the Compere-Song-Strominger~\cite{Compere:2013bya}(CSS) boundary conditions.

It turns out that the geodesic approximation will not work directly, and the matching in AdS$_3$/WCFT requires massive spinning particles in the bulk. As a byproduct, we also generalize the discussion of~\cite{Kraus:2016nwo} to operators with arbitrary spin in AdS$_3$/CFT$_2$. Within this generalization, the bulk picture in AdS$_3$/CFT and AdS$_3$/WCFT formally have the same tadpole structure Figure~\ref{plot1}, namely it contains a three point vertex, a bulk to boundary propagator $\langle\phi_{\mathcal O}\phi_{\mathcal O}\rangle$, and a propagator at the horizon $\langle\phi_\chi\phi_\chi\rangle$.
However, the holographic duality works in different ways, and the general formula is valid for different values of the mass and spin. In particular, the probe particle $\phi_{\mathcal O}$  in AdS$_3$/WCFT always has equal mass and spin, which cannot be described in the semiclassical approximation in AdS$_3$/CFT$_2$.

This paper is organized as follows. In section \ref{sec2} we compute a tadpole diagram in BTZ black holes for probes with arbitrary masses and spins. In section \ref{sec3} we calculate the average three-point coefficient in CFT$_2$, and match it with the bulk amplitude in the context of AdS$_3$/CFT$_2$. This part generalizes the discussion of~\cite{Kraus:2016nwo} which only considered scalars to operators with arbitrary spins in AdS$_3$/CFT$_2$.
In section \ref{sec4} and \ref{sec5}, we carry out a parallel analysis in AdS$_3$/WCFT. Using modular transformation to one-point functions on the torus, we derive an asymptotic formula of the average three-point coefficient for WCFT. Furthermore we match the result to a bulk tadpole diagram discussed in section  \ref{sec2}, using the holographic dictionary of AdS$_3$/WCFT.

\section{A tadpole diagram on BTZ black holes}\label{sec2}
In the quest for quantum gravity, many lessons have been learned from the study of three dimensional gravity. Under the Brown-Henneaux boundary conditions, Einstein gravity on AdS$_3$ with a negative cosmological constant is holographically dual to a two dimensional CFT. Under the CSS boundary conditions~\cite{Compere:2013bya}, the holographic dual is conjectured to be a WCFT, which is relevant for extremal black holes in higher dimensions.
Recently a new entry of holographic dictionary in AdS$_3$/CFT$_2$ has been found~\cite{Kraus:2016nwo}, that is,  the asymptotic growth of average three-point coefficient for scalar operators match a tadpole diagram in BTZ black holes in some parameter regions.  In the semiclassical limit, the bulk process can be approximated by point massive particles and hence the amplitude can be obtained from the geodesics.
In this section we revisit the same bulk process for generic massive spinning particles, generalizing the result of~\cite{Kraus:2016nwo} for scalars. In the following two sections, we will reproduce this bulk amplitude from either CFT or WCFT, respectively.
\begin{figure}[htbp]
\centering
\includegraphics[width=0.5\textwidth]{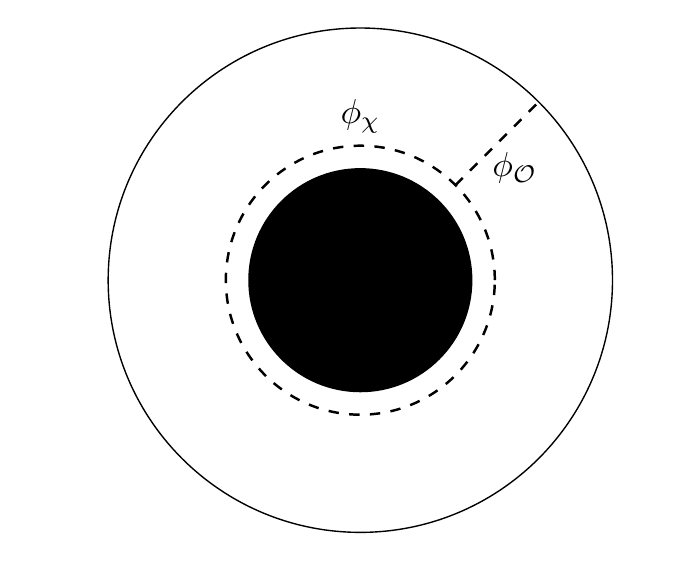}
\caption{ A massive spinning particle $\phi_{\mathcal{O}}$  propagates from infinity to horizon, where it splits into a pair of particles $\phi_{\chi}$ and $\phi_{\chi}^{\dagger}$ which wrap the horizon.}
\label{plot1}
\end{figure}

In the semiclassical limit, the bulk process \footnote{As was discussed in~\cite{Kraus:2017ezw}, to reproduce the contributions from all the descendent states of $|\chi\rangle$, a more careful geodesic Witten diagram~\cite{Hijano:2015zsa} should be considered. In that case, $\phi_\chi$ will be allowed to propagate in the entire BTZ spacetime, and the interaction point will be determined by minimizing the total worldline action. As a consequence, the one-point block in the bulk will be deformed. However, in this note, we will only focus on the contribution from the primary field and check how the picture works in the context of AdS$_3$/WCFT, as well as for arbitrary spins in AdS/CFT.} 
is as depicted in Figure~\ref{plot1}:
a massive spinning particle $\phi_{\mathcal{O}}$  propagates from infinity to horizon, where it splits into a pair of massive spinning particles $\phi_{\chi}$ and $\phi_{\chi}^{\dagger}$ which wrap the horizon. In the bulk theory, the two bulk field $\phi_{\mathcal{O}}$ and $\phi_{\chi}$ have a cubic interaction $\phi_{\mathcal O}\phi_{\chi}^\dagger \phi_
\chi$ proportional to the three point coefficient $\langle \chi|\mathcal{O}|\chi\rangle$ in the dual field theory. This amplitude $\mathcal{C}^{bk}$ is the product of the cubic-interaction vertex, the propagator of $\phi_{\mathcal{O}}$ from the boundary to the bulk, and the propagator of $\phi_{\chi}$ around the horizon. Schematically, we have $\mathcal{C}^{bk}\sim \langle \chi|\mathcal{O}|\chi\rangle \langle \phi_{\mathcal{O}}\phi_{\mathcal{O}}\rangle \langle \phi_\chi^\dagger\phi_\chi\rangle.$ In the following we will calculate the two propagators.

\subsection{Massive spinning particles in BTZ black holes}\label{subsec2.1}
In this subsection we describe the propagation of a massive spinning particles in BTZ spacetime. The metric of the BTZ black hole can be written in a form with light-like coordinates,
\begin{equation}\label{BTZ}
ds^2=\ell^2\left(T_u^2du^2+2\rho dudv+T_v^2dv^2+\frac{d\rho^2}{4(\rho^2-T_u^2T_v^2)}\right)\,,
\end{equation}
and identifications
\begin{equation}\label{id}
u\sim u+2\pi,~~~~v\sim v+2\pi\,.
\end{equation}
$T_u$ and $T_v$ are variable constants. The local isometry for the BTZ black hole is $SL(2, R)\times SL(2, R)$, while only the $U(1)\times U(1)$ part is globally well defined due to the spatial circle. It is useful to label the coordinates $u=\varphi+t$ and $v=\varphi-t$ where $t$ is the time and $\varphi$ is the space.

In the semiclassical approximation, a bulk two-point function for a field with mass $m$ and spin $s$ is given by $e^{-S}$, where $S$ is the on-shell worldline action of a spinning particle~\cite{Castro:2014tta},
\begin{equation}\label{worldlineaccft}
S=\int d\tau\left(m\sqrt{g_{\mu\nu}\dot{X}^{\mu}\dot{X}^{\nu}}+s\tilde{n}\cdot\nabla n\right)+S_{constraints}\,.
\end{equation}
Here $\tau$ is the length parameter. $S_{constraints}$ contains Lagrange multipliers which require that the two normalized  vector fields $n$ and $\tilde{n}$ should be mutually orthogonal and perpendicular to the worldline, namely
\begin{equation}\label{nnt}
n^2=-1,~~~~\tilde{n}^2=1,~~~~n\cdot\tilde{n}=0,~~~~n\cdot\dot{X}=\tilde{n}\cdot\dot{X}=0\,.
\end{equation}
The symbol $\nabla$ with no subscript indicates a covariant derivative alone the worldline:
\begin{equation}
\nabla V^{\mu}:=\frac{dX^{\nu}}{d\tau}\nabla_{\nu}V^{\mu}\,.
\end{equation}
The equations of motion with respect to $X^{\mu}(\tau)$ are known as the Mathisson-Papapetrou-Dixon (MPD) equations,
\begin{equation}
\nabla[m\dot{X}^{\mu}+\dot{X}^{\nu}\nabla s^{\mu}{}_{\nu}]=-\frac{1}{2}\dot{X}^{\nu}s^{\rho\sigma}R^{\mu}{}_{\nu\rho\sigma}\,,
\end{equation}
where $s^{\mu\nu}$ is the spin tensor,
\begin{equation}
s^{\mu\nu}=s(n^{\mu}\tilde{n}^{\nu}-\tilde{n}^{\mu}n^{\nu})\,.
\end{equation}
As was noticed in~\cite{Castro:2014tta}, in locally AdS spacetimes, the contraction of the Riemann tensor with $s^{\mu\nu}\dot{X}^{\rho}$ vanishes. The MPD equations reduce to
\begin{equation}\label{MPD0}
\nabla[m\dot{X}^{\mu}-s^{\mu}{}_{\nu}\nabla\dot{X}^{\nu}]=0\,.
\end{equation}
One obvious solution to the MPD equation above is a geodesic
\begin{equation}\label{geod}
\nabla\dot{X}^{\mu}=0\,.
\end{equation}
While other solutions to \eqref{MPD0} also exist, we will focus on the geodesic solution in this paper.

\subsection{A propagator from boundary to bulk}\label{subsec2.2}
For the particle $\phi_{\mathcal{O}}$ propagating from boundary to horizon along a radial geodesic, the on-shell worldline action (\ref{worldlineaccft}) has two parts. The first part is the length of the worldline times its mass. By using the argument above, a radial geodesic from infinity to the horizon can be viewed as the trajectory followed by $\phi_{\mathcal{O}}$, and its length with a cutoff $\Lambda$ can be evaluated through the metric (\ref{BTZ}),
\begin{equation}\label{length}
L_{\mathcal{O}}=\ell\int_{T_uT_v}^{\Lambda}\frac{d\rho}{2\sqrt{\rho^2-T_u^2T_v^2}}\approx\frac{\ell}{2}(\log2\Lambda-\log T_uT_v)+\cdots\,.
\end{equation}

The second part in (\ref{worldlineaccft}) comes from the spin contribution, which we denote as $S_{\mathrm{spin}}$. As discussed in~\cite{Castro:2014tta}, it can be shown that this term only depends on the boundary data of the normal vectors $n(\tau)$ or $\tilde{n}(\tau)$. Explicitly,
\begin{equation}
S_{\mathrm{spin}}=s_{\mathcal{O}}\log\left(\frac{q(\tau_f)\cdot n_f-\tilde{q}(\tau_f)\cdot n_f}{q(\tau_i)\cdot n_i-\tilde{q}(\tau_i)\cdot n_i}\right)\,,
\end{equation}
where two vectors $q^{\mu}$ and $\tilde{q}^{\mu}$ are mutually orthogonal, perpendicular to the geodesic, and furthermore are parallel transported along and the geodesic, i.e., \bea q^2=-1,\quad \tilde{q}^2=1,\quad q\cdot\tilde{q}=0,\quad q\cdot\dot{X}=\tilde{q}\cdot\dot{X}=0,\quad \nabla q=\nabla\tilde{q}=0.  \eea
The two sets of vectors $(n(\tau), \tilde{n}(\tau))$ and $(q(\tau), \tilde{q}(\tau))$ can be related via a Lorentz boost. In fact, we can expand $n(\tau)$ and $\tilde{n}(\tau)$ in terms of $q(\tau)$ and $\tilde{q}(\tau)$,
\begin{eqnarray}
n(\tau)&=&\cosh(\eta(\tau))q(\tau)+\sinh(\eta(\tau))\tilde{q}(\tau)\,,\\
\tilde{n}(\tau)&=&\sinh(\eta(\tau))q(\tau)+\cosh(\eta(\tau))\tilde{q}(\tau)\,,
\end{eqnarray}
where $\eta(\tau)$ is the rapidity of this Lorentz boost. $S_{\mathrm{spin}}$ measures the total change of this boost and can also be expressed in terms of $\tilde{n}$, $q$, and $\tilde{q}$,
\begin{equation}\label{Sanom}
S_{\mathrm{spin}}=s_{\mathcal{O}}\log\left(\frac{\tilde{q}(\tau_f)\cdot\tilde{n}_f-q(\tau_f)\cdot\tilde{n}_f}{\tilde{q}(\tau_i)\cdot\tilde{n}_i-q(\tau_i)\cdot\tilde{n}_i}\right)\,.
\end{equation}
Using the metric (\ref{BTZ}), it is straightforward to find the tangent vector and a parallel transport normal frame of the radial geodesic,
\begin{eqnarray}
\dot{X}^{\mu}\partial_{\mu}&=&\frac{2}{\ell}\sqrt{\rho^2-T_u^2T_v^2}\partial_{\rho},\\
\tilde{q}^{\mu}\partial_{\mu}&=&\frac{T_v}{\ell\sqrt{2T_uT_v(\rho+T_uT_v)}}\partial_u+\frac{T_u}{\ell\sqrt{2T_uT_v(\rho+T_uT_v)}}\partial_v\,,\nonumber\\
q^{\mu}\partial_{\mu}&=&\frac{T_v}{\ell\sqrt{2T_uT_v(\rho-T_uT_v)}}\partial_u-\frac{T_u}{\ell\sqrt{2T_uT_v(\rho-T_uT_v)}}\partial_v\,.
\end{eqnarray}
Setting the initial point at infinity and final point at the horizon, it is easy to find \footnote{Here we choose the time-like co-vector field $n_{\mu}dx^{\mu}$ to be proportional to $dt$ along the worldline, and the spatial normal co-vector $\tilde{n}_{\mu}dx^{\mu}$ which is perpendicular to $n$ always takes a form,
\begin{equation}
\tilde{n}_{\mu}dx^{\mu}\propto d\varphi+\frac{T_u^2-T_v^2}{2\rho+T_u^2+T_v^2}dt\,.\nonumber
\end{equation}
The worldline action (\ref{worldlineaccft}) is covariant, thus insensitive to the specific choices of $n$ and $\tilde{n}$.},
\begin{equation}
\tilde{n}_i^{\mu}\partial_{\mu}=\lim_{\rho\to\infty}\frac{1}{\ell\sqrt{2\rho}}(\partial_u+\partial_v),~~~~\tilde{n}_f^{\mu}\partial_{\mu}=\lim_{\rho\to T_uT_v}\frac{1}{\ell(\rho+T_uT_v)}(T_v\partial_u+T_u\partial_v)\,.
\end{equation}
Substituting the above formulas into (\ref{Sanom}), we find,
\begin{equation}
S_{\mathrm{spin}}=s_{\mathcal{O}}\log\sqrt{\frac{T_v}{T_u}}\,.
\end{equation}
Putting the two parts together, we can explicitly write down the regularized radial boundary to bulk propagator,
\begin{equation}\label{bhpo}
e^{-S_{\mathcal{O}}}=T_u^{\frac{\ell m_{\mathcal{O}}+s_{\mathcal{O}}}{2}}T_v^{\frac{\ell m_{\mathcal{O}}-s_{\mathcal{O}}}{2}}\,.
\end{equation}

\subsection{A propagator at the horizon}\label{subsec2.3}
In this subsection, we consider the propagator of a massive spinning particle $\phi_{\chi}$ going around the horizon as shown in the Figure~\ref{plot1}. Similar to the discussion for the particle $\phi_\mathcal {O}$, the amplitude for $\phi_{\chi}$ wrapping the horizon can be evaluated as $e^{-S_{\chi}}$ in the semi-classical limit. Again we will take the geodesic solution to the MPD equation. Then the total on-shell worldline action of $\phi_{\chi}$, analogous to \eqref{worldlineaccft}, will become
\bea \label{worldlineacchicft}S_{\chi}&=&m_\chi L_\chi+S_{\chi}^{\mathrm{spin}}\nonumber\\
&=&2\pi\ell m_\chi(T_u+T_v)  +s_{\chi} \int d\tau \tilde{n}\cdot\nabla n
\eea where the first part is the length of the trajectory at the horizon times the mass, and the second part is the spin contribution,
$m_{\chi}$ and $s_{\chi}$ are the mass and spin of $\phi_{\chi}$ respectively, $n^{\mu}$ and $\tilde{n}^{\mu}$ are mutually orthogonal normal vectors as defined in (\ref{nnt}). The trajectory of $\phi_{\chi}$ is the horizon at constant time $t$ and is also a geodesic satisfying (\ref{geod}).

The horizon is a degenerate hypersurface, and direct solution for $n$ and $\tilde {n}$ will be singular.  Instead, we take the orbit of the particle $\phi_{\chi}$ to be a circular orbit with constant radius slightly greater than the horizon's. It can be shown that the constant radius orbits outside the horizon are no longer geodesics, but with accelerations. Then we evaluate the action of $\phi_{\chi}$ on that orbit, this is not on-shell since (\ref{MPD0}) is not satisfied. Finally, we take the radius of that orbit tends to the horizon. On the horizon, the worldline action become on-shell, and we will get the result.
By using the metric (\ref{BTZ}), we can find out the tangent and normal vectors of the horizon at constant time followed by $\phi_{\chi}$,
\begin{eqnarray}
\dot{X}^{\mu}\partial_{\mu}&=&\lim_{\rho\to TuTv}\frac{1}{\ell\sqrt{2\rho+T_u^2+T_v^2}}\partial_u+\frac{1}{\ell\sqrt{2\rho+T_u^2+T_v^2}}\partial_v,\\ \tilde{n}^{\mu}\partial_{\mu}&=&\lim_{\rho\to TuTv}\frac{2\sqrt{\rho^2-T_u^2T_v^2}}{\ell}\partial_{\rho}\,,\\
n^{\mu}\partial_{\mu}&=&\lim_{\rho\to TuTv}\frac{\rho+T_v^2}{\ell\sqrt{(\rho^2-T_u^2T_v^2)(2\rho+T_u^2+T_v^2)}}\partial_u-\frac{\rho+T_u^2}{\ell\sqrt{(\rho^2-T_u^2T_v^2)(2\rho+T_u^2+T_v^2)}}\partial_v\,.\nonumber\\
\end{eqnarray}
Substituting the above expressions in to the action (\ref{worldlineacchicft}),  we learn that the limit $\rho\to T_uT_v$ is finite and furthermore the integrand of $S^{\mathrm{spin}}_\chi$ is independent of the affine parameter.  Therefore we can directly perform the integral at the horizon, and get $S^{\mathrm{spin}}_\chi=s_{\chi}2\pi(T_u-T_v)$. Alternatively, the same result can be obtained by the formula \eqref{Sanom}, similar to the discussion for $\mathcal{O}$. More details for the alternative derivation can be found in Appendix A.

Putting the two contributions together, the total on-shell action of the massive spinning particle $\phi_{\chi}$ going round the horizon once is given by
\begin{equation}\label{os}
S_{\chi}
=2\pi(\ell m_{\chi}+s_{\chi})T_u+2\pi(\ell m_{\chi}-s_{\chi})T_v\,.
\end{equation} Note that \eqref{os} is independent of boundary conditions or the holographic dictionaries, and will be used in the story of AdS$_3$/CFT$_2$ and AdS$_3$/WCFT in next sections.

\subsubsection*{Backreactions}
If the mass of the particle is comparable to ${1\over G}$, backreactions to the background geometry will be important if we try to match the local mass and spin in the bulk to quantum numbers in the dual field theory.
Solving the Einstein equation with a local source, the backreaction of a particle with mass $m_\chi$ and $s_\chi$ geometry will be a rotational conical defect~\cite{Deser:1983tn, Deser:1983nh},
\bea\label{conical}
ds^2&=&\ell^2\left[-(1+r^2)\left(dt-\frac{s_{\chi}}{\frac{\ell}{4G}\left(1-\frac{\delta\varphi}{2\pi}\right)}d\varphi\right)^2+\frac{dr^2}{1+r^2}+r^2d\varphi^2\right],~~~~\\
\varphi&\sim&\varphi+2\pi-\delta\varphi\,.
\eea
Here the deficit angle $\delta\varphi$ is related to the mass of the $\phi_{\chi}$ through
\begin{equation}\label{massx}
m_{\chi}=\frac{\delta\varphi}{8\pi G}\,.
\end{equation}
More details of the solution can be found in Appendix B, where we show that  the mass and spin of the local source can be obtained from the geometry by the quasi-local energy.

In order to match results in the boundary, we need to put the conical defect solution \eqref{conical} into the form of \eqref{BTZ} with standard identification \eqref{id}, from which we can read the asymptotic charges easily. One can verify that the coordinate transformation is
\begin{eqnarray}
u&=&\frac{\varphi}{1-4G m_\chi}+\frac{t}{1-4G(m_\chi+s_{\chi}/\ell)}\,,\\
v&=&\frac{\varphi}{1-4G m_\chi}-\frac{t}{1-4G(m_\chi-s_{\chi}/\ell)}\,,\\
\rho&=&\frac{1}{2}\left[\left(1-4G m_\chi\right)^2-16G^2s_{\chi}^2/\ell^2\right](\frac{1}{2}+r^2)\,.
\end{eqnarray}
The resulting metric has fixed boundary metric and identifications \eqref{id}, and takes the standard form,
\bea\label{brg}
ds^2&=&\ell^2\left[T_{\chi u}^2du^2+2\rho dudv+T_{\chi v}^2dv^2+\frac{d\rho^2}{4(\rho^2-T_{\chi u}^2T_{\chi v}^2)}\right]\,,\\
(u,v)&\sim& (u+2\pi,\, v+2\pi)\nonumber
\eea
where
\begin{equation}\label{TxuTxv}
T_{\chi u}^2=-\frac{(1-4G(m_\chi+s_{\chi}/\ell))^2}{4},~~~~T_{\chi v}^2=-\frac{(1-4G(m_\chi-s_{\chi}/\ell))^2}{4}\,.
\end{equation}
Note that \eqref{brg} is in both the Brown-Henneaux phase space and the CSS phase space. From \eqref{brg}, we can easily read off the asymptotic charges, and are related to quantum numbers in the field theory according to the holographic dictionary.

\subsection{A tadpole diagram in the bulk}\label{subsec2.4}
Combing the two contributions from $\phi_{\mathcal{O}}$ (\ref{bhpo}) and $\phi_{\chi} $(\ref{os}), we get the total amplitude for the process given by Figure~\ref{plot1},
\begin{equation}\label{cbk}
\mathcal{C}_{\mathcal{O}}^{bk}( E_L , {E_R})=\underbrace{\langle \chi|\mathcal{O}|\chi\rangle}_{\text{ vertex}} \,\underbrace{T_u^{\frac{\ell m_{\mathcal{O}}+s_{\mathcal{O}}}{2}}T_v^{\frac{\ell m_{\mathcal{O}}-s_{\mathcal{O}}}{2}}}_{\langle  \phi_{\mathcal{O}}  \phi_{\mathcal{O}}\rangle}\,\underbrace{e^{-\ell m_{\chi}2\pi(T_v+T_u)-s_{\chi}2\pi(T_u-T_v)}}_{\langle  \phi_{\chi}^\dagger\phi_{\chi}\rangle}\,,
\end{equation}
which describes a massive spinning particle $\phi_{\mathcal{O}}$ propagates from infinity, and decays into a loop of $\phi_\chi$ around the black hole horizon.

\section{Structure constant in AdS$_3$/CFT$_2$ with spinning operators}\label{sec3}
In this section, we consider the mean value of three-point coefficients in the context of AdS$_3$/CFT$_2$. Our results generalize those of~\cite{Kraus:2016nwo} to operators with arbitrary spins.
On the CFT$_2$ side, we derive the typical expectation value of an operator $\mathcal{O}$ with conformal weights $(h_{\mathcal{O}}, \bar{h}_{\mathcal{O}})$ on states with weights $( E_L +{c\over24}, E_R +{c\over24})$.  
We follow the strategy of~\cite{Kraus:2016nwo} where the same quantity was calculated for scalar operator $\mathcal{O}$.
By changing the ensemble, the typical value of three-point function can be calculated from the torus one-point function.
Similar to the derivation of Cardy's formula,
modular invariance of two dimensional CFTs allows an estimation of torus one-point function at high temperature, or equivalently at high energy. On the gravity side, we consider Einstein gravity in asymptotic AdS$_3$ spacetime with the Brown-Henneaux boundary conditions.
The gravity picture is then a tadpole diagram on BTZ background as depicted in Figure~\ref{plot1}. Due to the spin, the propagator in the bulk will be given by the worldline action of a spinning particle instead of a spinless particle as discussed in~\cite{Kraus:2016nwo}.
In the following we will show the CFT$_2$ calculation and bulk calculation respectively.

\subsection{Asymptotic structure constant in CFT$_2$ with arbitrary spins}\label{subsec3.1}
In this subsection, we derive the mean structure constant of an operator $\mathcal{O}$ with conformal weights $(h_{\mathcal{O}}, \bar{h}_{\mathcal{O}})$ and two operators with conformal weights $( E_L +{c\over24}, E_R +{c\over24})$ in the limit of $ E_L ,{E_R}  \gg1. $ 
In the following, we will carry out a detailed calculation in CFT$_2$, for the purpose of easy comparison with that of WCFT in the next section.

In a two dimensional CFT, the one-point function of a primary operator $\mathcal{O}$ of conformal weights $(h_{\mathcal{O}}, \bar{h}_{\mathcal{O}})$ on a torus with complex modular parameter $\tau=i\tau_1+\tau_2$ can be written in terms of three-point coefficients as follows,
\be
\langle\mathcal{O}\rangle_{\tau, \bar{\tau}}=\mathrm{Tr}\mathcal{O}e^{2\pi i\tau L_0-2\pi i\bar{\tau}\bar{L}_0}=\int d E_L  d{E_R}T_{\mathcal{O}}( E_L , {E_R})e^{2\pi i\tau E_L -2\pi i\bar{\tau}{E_R}}\,,\label{onetothree}
\ee
where
\be
T_{\mathcal{O}}( E_L , {E_R})= \sum_i \langle i|\mathcal{O}|i\rangle\delta( E_L - {E_L} _i)\delta( E_R -{ E_R }_i)\,.
\ee
Up to normalizations, $\langle i|\mathcal{O}|i\rangle$ is the OPE coefficient $c_{ii\mathcal{O}}$, and hence $ T_{\mathcal{O}}( E_L , {E_R})$ is the total three-point coefficient from all operators with $( {E_L}_i, {E_R}_i)$. $L_0$ and $\bar{L}_0$ are the zero modes of the standard Virasoro algebra.
$ E_L $ and ${E_R}$ are the eigenvalues of $L_0$ and $\bar{L}_0$ on the cylinder, respectively, with relative shifts $ E_L  ({E_R} )=h(\bar h) -\frac{c}{24}$ to the conformal weights  $(h,\bar h)$ defined on the plane. The imaginary part of $\tau$ is proportional to the inverse temperature and the real part is the angular potential. Here we keep them arbitrary, and treat $\tau$ and $\bar{\tau}$ as independent complex variables.

We can invert  the relation \eqref{onetothree} between the torus one-point function and $T_{\mathcal{O}}( E_L , {E_R})$ by an  inverse Laplace transformation.  From this perspective,
the total three-point coefficient $T_{\mathcal{O}}( E_L , {E_R})$ is also the density of states weighted by the one-point function,
\begin{equation}\label{Tdd}
T_{\mathcal{O}}( E_L , {E_R})=\int d\tau d\bar{\tau}e^{-2\pi i\tau E_L }e^{2\pi i\bar{\tau}{E_R}}\langle\mathcal{O}\rangle_{\tau, \bar{\tau}}\,.
\end{equation}
Two dimensional CFTs are invariant under the large conformal transformations on the torus, which act on $\tau$ as modular transformations,
\begin{equation}
\tau\to\gamma\tau\equiv\frac{a\tau+b}{c\tau+d}\,,
\end{equation}
where $ad-cd=1$.
The partition function of theory is invariant under such modular transformation, and the primary operator $\mathcal{O}$ transforms with the modular weights $(h_{\mathcal{O}}, \bar{h}_{\mathcal{O}})$,
\begin{equation}
\langle\mathcal{O}\rangle_{\gamma\tau, \gamma\bar{\tau}}=(c\tau+d)^{h_{\mathcal{O}}}(c\bar{\tau}+d)^{\bar{h}_{\mathcal{O}}}\langle\mathcal{O}\rangle_{\tau, \bar{\tau}}\,.
\end{equation}
In particular, under the $S$ transformation $\tau\to-1/\tau$, the one-point function transforms as
\begin{equation}\label{modO}
\langle\mathcal{O}\rangle_{\tau, \bar{\tau}}=\tau^{-h_{\mathcal{O}}}\bar{\tau}^{-\bar{h}_{\mathcal{O}}}\langle\mathcal{O}\rangle_{-1/\tau, -1/\bar{\tau}}\,.
\end{equation}
This formula is very useful as it relates the high temperature behaviour of the theory to the behaviour at low temperature. Taking the limit $\tau\to i0^+$, $\bar{\tau}\to-i0^+$, and assuming the eigenvalues of $L_0$ and $\bar{L}_0$ are bounded from below, we can project the right hand side of (\ref{modO}) onto a lightest state $|\chi\rangle$ with non-vanishing three-point coefficient $\langle\chi|\mathcal{O}|\chi\rangle\neq0$ \footnote{Throughout this paper, we focus on the contribution from the primary field $\chi$, without considering the possible degeneracy.  When $\chi$ has degeneracies, we need to replace $\langle\chi|\mathcal{O}|\chi\rangle$ by a sum over all degenerate states.  Subtleties may appear when the sum is zero. Related discussions can be found in~\cite{Kraus:2016nwo}. Here we will only consider the most general theories where such coincidence doesn't show up. See~\cite{Kraus:2017ezw} for the contributions from all the descendants.},
\begin{equation}\label{Ot}
\langle\mathcal{O}\rangle_{\tau, \bar{\tau}}=\langle\chi|\mathcal{O}|\chi\rangle\tau^{-h_{\mathcal{O}}}\bar{\tau}^{-\bar{h}_{\mathcal{O}}}e^{-2\pi i\frac{1}{\tau} {E_L}_{\chi}}e^{2\pi i\frac{1}{\bar{\tau}}{E_R}_{\chi}}\,,
\end{equation}
where $( {E_L}_{\chi}, {E_R}_{\chi})$ are the conformal dimensions of the state $|\chi\rangle$. Substituting (\ref{Ot}) back into (\ref{Tdd}), we can write down the total three-point coefficient in terms of $ {E_L}_{\chi}$ and ${E_R}_{\chi}$,
\begin{equation}\label{Tddx}
T_{\mathcal{O}}( E_L , {E_R})=\langle\chi|\mathcal{O}|\chi\rangle\int d\tau d\bar{\tau}\tau^{-h_{\mathcal{O}}}\bar{\tau}^{-\bar{h}_{\mathcal{O}}}e^{-2\pi i\tau E_L }e^{2\pi i\bar{\tau}{E_R}}e^{-2\pi i\frac{1}{\tau} {E_L}_{\chi}}e^{2\pi i\frac{1}{\bar{\tau}}{E_R}_{\chi}}\,.
\end{equation}
At large $ E_L $ and ${E_R}$, the above integral is dominated by a saddle point with
\begin{equation}
\tau=i\sqrt{-\frac{ {E_L}_{\chi}}{ E_L }},~~~~\bar{\tau}=-i\sqrt{-\frac{{E_R}_{\chi}}{{E_R}}}\,.
\end{equation}
Further assuming $ {E_L}_\chi,\,{ E_R }_\chi <0$, the saddle points are pure imaginary.
Under the saddle point approximation, the integral in (\ref{Tddx}) can be performed and we get,
\begin{equation}\label{Tddxs}
T_{\mathcal{O}}( E_L , {E_R})=i^{\bar{h}_{\mathcal{O}}-h_{\mathcal{O}}}\langle\chi|\mathcal{O}|\chi\rangle\left(-\frac{ {E_L}_{\chi}}{ E_L }\right)^{-\frac{h_{\mathcal{O}}}{2}}
\left(-\frac{{E_R}_{\chi}}{{E_R}}\right)^{-\frac{\bar{h}_{\mathcal{O}}}{2}}e^{4\pi\sqrt{- E_L  {E_L}_{\chi}}+4\pi\sqrt{-{E_R}{E_R}_{\chi}}}\,.
\end{equation}
$T_{\mathcal{O}}( E_L , {E_R})$ characterizes the total contribution of different degenerate states to the three-point function coefficient of the underling CFT at given large $ E_L $ and ${E_R}$. However, it is useful to define the typical value of the three-point coefficient by dividing $T_{\mathcal{O}}( E_L , {E_R})$ by the density of states. Choosing the operator $\mathcal{O}$ being the identity and setting the state $|\chi\rangle$ to the vacuum in (\ref{Tddxs}), we can write down the density of states $\rho( E_L , {E_R})$ at large $ E_L $ and ${E_R}$,
\begin{equation}
\rho( E_L , {E_R})=e^{4\pi\sqrt{- E_L  {E_L}_{vac}}+4\pi\sqrt{-{E_R}{E_R}_{vac}}}\,.
\end{equation}
Taking $ {E_L}_{vac}={E_R}_{vac}=-\frac{c}{24}$, the above equation is nothing but the Cardy's formula in two dimensional CFT. The mean structure constant for the operator $\mathcal{O}$ and two operators with same large conformal dimension $ E_L $ and ${E_R}$ can be defined,
\bea\label{sccft}
&&\mathcal{C}_{\mathcal{O}}( E_L , {E_R})\equiv\frac{T_{\mathcal{O}}( E_L , {E_R})}{\rho( E_L , {E_R})}\\ 
&&=i^{\bar{h}_{\mathcal{O}}-h_{\mathcal{O}}}\langle\chi|\mathcal{O}|\chi\rangle\left(-\frac{ {E_L}_{\chi}}{ E_L }\right)^{-\frac{h_{\mathcal{O}}}{2}}
\left(-\frac{{E_R}_{\chi}}{{E_R}}\right)^{-\frac{\bar{h}_{\mathcal{O}}}{2}}e^{4\pi\sqrt{ E_L }(\sqrt{- {E_L}_{\chi}}-\sqrt{- {E_L}_{vac}})+4\pi\sqrt{{E_R}}(\sqrt{-{E_R}_{\chi}}-\sqrt{-{E_R}_{vac}})}\,.\nonumber
\end{eqnarray}
This is the asymptotic formula of the average value of three-point coefficient for a general CFT$_2$, assuming $ E_L ,{E_R} \gg 1,$ and the existence of an operator $\chi$ with $ {E_L}_\chi,\,{ E_R }_\chi <0$ and  $\langle\chi|\mathcal{O}|\chi\rangle\neq0$. 
When $\mathcal O$ is a scalar operator with $h_{\mathcal O}={\bar h}_{\mathcal O}$, \eqref{sccft} reduces to the CFT results of~\cite{Kraus:2016nwo}.
In the following subsection, we will match the bulk calculation \eqref{cbk} to the CFT result \eqref{sccft}.

\subsection{Matching with AdS$_3$/CFT$_2$}\label{subsec3.2}
Under Brown-Henneaux boundary conditions, asymptotically AdS$_3$ spacetimes are dual to CFT$_2$s.
In the context of AdS$_3$/CFT$_2$,~\cite{Kraus:2016nwo} proposed a bulk dual for the asymptotic formula of mean three-point coefficient for scalar operators in the heavy-heavy-light limit.
In this subsection, we check the picture for the generic spins by
matching the bulk tadpole diagram  with the CFT quantity $\mathcal{C}_{\mathcal{O}}( E_L , {E_R})$~(\ref{sccft}).
In the following, we will first use the AdS$_3$/CFT$_2$ dictionary to rewrite the result (\ref{sccft}) in terms of BTZ parameters, and then consider the contributions from $\phi_{\mathcal O}$ and $\phi_{\chi}$, and finally match the bulk result \eqref{cbk} to \eqref{sccft}.

We consider the average value of the three-point coefficient $\mathcal{C}_{\mathcal{O}}( E_L , {E_R})$ (\ref{sccft}), in the heavy-heavy-light limit with  $1\ll h_{\mathcal{O}}, \bar{h}_{\mathcal{O}}\ll\frac{c}{24}$ and  $ E_L ,\,  E_R \ge {c\over24},\, c\gg1.$  At large $ E_L $ and ${E_R}$, a typical state $| E_L , {E_R}\rangle$ is well described by the black hole geometry in AdS$_3$, which emerges after coarse-graining over many states at fixed $ E_L $ and ${E_R}$. The difference from ~\cite{Kraus:2016nwo} is that for $ E_L \neq{E_R}$, we will need to consider rotating BTZ black holes.
More explicitly, the quantum numbers $ E_L $ and $ E_R $ in CFT$_2$ are the conserved charges associate with the Killing vectors $\partial_u$ and $-\partial_v$ evaluated on the BTZ metric (\ref{BTZ}),
\begin{equation}\label{ddb}
 E_L =Q[\partial_u]=\frac{\ell T_u^2}{4G},~~~~{E_R}=Q[-\partial_v]=\frac{\ell T_v^2}{4G}\,.
\end{equation}
The dual geometry for the CFT vacuum is the global AdS$_3$, which corresponds to $T_{u, v}=-\frac{i}{2}$, with eigenvalues
\begin{equation}\label{dvdbv}
 {E_L}_{vac}={E_R}_{vac}=-{\ell\over 16G}\,.
\end{equation}

For a light operator $\mathcal{O}$ with $1\ll h_{\mathcal{O}}, \bar{h}_{\mathcal{O}}\ll\frac{c}{24}$,  its dual field $\phi_{\mathcal{O}}$ can be  approximated by a massive spinning particle, instead of a spinless particle when  $h_{\mathcal{O}}={\bar h}_{\mathcal{O}}$ as considered in~\cite{Kraus:2016nwo}.
Under the massive particle approximation, the conformal weights for $\mathcal{O}$ are related to the mass and spin by
\be
h_{\mathcal{O}}=\frac{1}{2}(\ell m_{\mathcal O}+s_{\mathcal{O}}),~~~~\bar{h}_{\mathcal{O}}=\frac{1}{2}(\ell m_{\mathcal O}-s_{\mathcal{O}})\,.\label{oms}
\ee
Therefore the propagator for $\phi_{\mathcal{O}}$ \eqref{bhpo} can be also written as
\begin{equation}\label{bhp1}
e^{-S_{\mathcal{O}}}=T_u^{h_{\mathcal O}}T_v^{{\bar h}_{\mathcal O} }\,.
\end{equation}

We will further take the lightest state $|\chi\rangle$ coupled to  $\mathcal{O}$ to be a heavy operator in the region $1\ll h_{\chi}, \bar{h}_{\chi}<\frac{c}{24}$. Then the dual bulk field, denoted by $\phi_\chi$, will correspond to a non-perturbative field which backreacts on the bulk AdS$_3$ geometry but still below the black hole threshold. This means that $\phi_\chi$ will create a spinning conical defect for generic $h_\chi-{\bar h}_\chi$.
Now we need to relate the local bulk mass $m_\chi$ and spin $s_\chi$ to conformal weights $h_\chi,\,{\bar h}_\chi$ in the dual CFT$_2$. As described in section 2.3, the spinning conical defect \eqref{conical} can be brought into a standard form \eqref{brg} in the Brown-Henneax phase space.
Then the conformal weights can be read from the asymptotic charges evaluated on \eqref{brg} with the parameters \eqref{TxuTxv},
\bea
 {E_L}_\chi=Q_\chi[\partial_u]=\frac{\ell T_{\chi u}^2}{4G},~~~~{E_R}_\chi=Q_\chi[-\partial_v]=\frac{\ell T_{_\chi v}^2}{4G}
\eea
Using the above map and Eqs. (\ref{TxuTxv}), one can rewrite the mass and spin of $\phi_{\chi}$ in terms of its conformal dimensions,
\begin{equation}\label{mchischi}
\ell m_{\chi}=\frac{\ell}{4G}-\sqrt{-\frac{\ell {E_L}_{\chi}}{4G}}-\sqrt{-\frac{\ell{E_R}_{\chi}}{4G}},~~~~s_{\chi}=-\sqrt{-\frac{\ell {E_L}_{\chi}}{4G}}+\sqrt{-\frac{\ell{E_R}_{\chi}}{4G}}\,.
\end{equation}
Substituting the mass and spin formula above into (\ref{os}), the amplitude for $\phi_{\chi}$ wrapping the horizon can be recast as
\begin{equation}\label{bhp2}
e^{-S_{\chi}}=e^{4\pi\left[T_u\left(\sqrt{-\frac{\ell {E_L}_{\chi}}{4G}}-\frac{\ell}{8G}\right)+T_v\left(\sqrt{-\frac{\ell{E_R}_{\chi}}{4G}}-\frac{\ell}{8G}\right)\right]}\,.
\end{equation}
Putting the two parts \eqref{bhp1} and \eqref{bhp2} together,  the bulk amplitude \eqref{cbk} for the process given by Figure~\ref{plot1} under Brown-Henneaux boundary conditions can be written as
\bea
\mathcal{C}^{bk}_{\mathcal{O}}( E_L ,{E_R} )&=&\underbrace{<\chi|\mathcal{O}|\chi>}_{\text{ vertex}}\,\underbrace{T_u^{h_{\mathcal O}}T_v^{{\bar h}_{\mathcal O}}}_{\langle  \phi_{\mathcal{O}}  \phi_{\mathcal{O}}\rangle}\,
\underbrace{ e^{4\pi\left[T_u\left(\sqrt{-\frac{\ell {E_L}_{\chi}}{4G}}-\frac{\ell}{8G}\right)+T_v\left(\sqrt{-\frac{\ell{E_R}_{\chi}}{4G}}-\frac{\ell}{8G}\right)\right]}}_{\langle  \phi_{\chi} \phi_{\chi}\rangle}\, \label{bkAdSCFT}
\eea
Up to an overall normalization that does not depend on $ E_L $ and ${E_R} $, the above bulk amplitude agrees with the average value of three-point coefficient  (\ref{sccft})  rewritten in terms of temperatures.
Note that
the bulk approximation \eqref{bkAdSCFT} in the  AdS$_3$/CFT$_2$ correspondence is valid in the region $1\ll h_{\mathcal{O}},\,\bar h_{\mathcal{O}}\ll {c\over24},$ and $1\ll h_{\chi}, \bar{h}_{\chi}<\frac{c}{24}$.
In particular, \eqref{bkAdSCFT} is not valid for $\bar h_{\mathcal{O}}=0$ which corresponds to particles with equal mass and spin in the bulk. This is in contrast with the WCFT result \eqref{twoparts} in the next section which is only valid for $\phi_{\mathcal O}$ with equal mass and spin.

\section{The AdS$_3$/WCFT correspondence}\label{sec4}
\subsection{Warped CFT and modular invariance}\label{subsec4.1}
In this subsection, we briefly review some basic features of WCFT.
A warped conformal field theory is characterized by the warped conformal symmetry. The global symmetry is $SL(2, R)\times U(1)$ , while the local symmetry algebra is a Virasoro algebra plus a $U(1)$ Kac-Moody algebra~\cite{Hofman:2011zj, Detournay:2012pc}. In position space, a general warped conformal symmetry transformation can be written as
\begin{equation}
x'=f(x),~~~~y'=y+g(x)\,,
\end{equation}
where $x$ and $y$ are $SL(2, R)$ and $U(1)$ local coordinates, and $f(x)$ and $g(x)$ are two arbitrary functions. Denote $T(x)$ and $P(x)$ as the Noether currents associated with the translation along $x$ and $y$ axis, respectively. The commutation relations for the Noether charges form a canonical warped conformal algebra consists of one Virasoro algebra and a Kac-Moody algebra,
\begin{eqnarray}\label{calg}
{}[L_n, L_m]&=&(n-m)L_{n+m}+\frac{c}{12}n(n^2-1)\delta_{n,-m},\nonumber\\
{}[L_n, P_m]&=&-mP_{n+m},\\
{}[P_n, P_m]&=&n\frac{k}{2}\delta_{n,-m}\,,\nonumber
\end{eqnarray}
where $c$ is the central charge and $k$ is the Kac-Moody level. One can also construct the spectral flow invariant Virasoro generators,
\bea\label{spfinvL}
L^{inv}_{n}&=&L_n-{1\over k} \Big(\sum_{m\le-1}P_{m}P_{n-m}+\sum_{m\ge0}P_{n-m}P_m\Big)\,.
\eea
One can check that $L^{inv}_{n}$ commutate with the Kac-Moody generators, and
form a Virasoro algebra with central charge $c-1$.

Now consider a WCFT with coordinates $(x, y)$ on a torus with two circles
\bea\label{spi}
spatial \, circle: &&\quad (x,y)\sim (x+2\pi,y)\,,
\\
\label{thermalid}
thermal \, circle:&&\quad (x, y)\sim(x+i\beta, y-i\bar{\beta})\,,
\eea
where $\beta$ and $\bar{\beta}$ are the inverse temperatures along $x$ and $y$, respectively. The torus partition function can be written as \begin{equation}\label{PF}
Z(\beta, \bar{\beta})=\mathrm{Tr}(e^{-\beta L_0+\bar{\beta}P_0})\,,
\end{equation}
where $L_0$ and $P_0$ are the zero modes of the generators, which are the conserved charges associated with the coordinates $x$ and $y$ respectively,
\begin{equation}
L_0=Q[\partial_x],~~~~P_0=Q[\partial_y]\,,
\end{equation}
The partition function \eqref{PF} transforms covariantly under modular transformation, as first obtained in~\cite{Detournay:2012pc}, and further explained in~\cite{Castro:2015csg}.
First note that the partition function is invariant under swapping the circles to
\bea spatial\, circle:&&\quad (x,y)\sim(x+i\beta,y-i\bar\beta)\,,\\
thermal\, circle:&&\quad (x,y)\sim(x-2\pi,y)\,.\eea
Then the following warped conformal transformation
\begin{equation}\label{St}
x'=-i\frac{2\pi}{\beta} x,~~~~y'=y+\frac{\bar{\beta}}{\beta} x\,,
\end{equation}
leads to a torus with new circles,
\bea
spatial\, circle:&&\quad  (x', y')\sim(x'+2\pi, y')\,,\\
thermal\, circle:&&\quad (x', y')\sim(x'+i\beta', y'-i\bar{\beta}')\,,
\eea
where
\begin{equation}
\beta'=\frac{4\pi^2}{\beta},~~~~\bar{\beta}'=-\frac{2\pi i\bar{\beta}}{\beta}\,.
\end{equation}
The partition function transforms according to the following equation with anomaly~\cite{Detournay:2012pc, Castro:2015csg},
\bea
Z(\beta, \bar{\beta})
&=&e^{k\frac{\bar{\beta}^2}{4\beta}}Z\left(\frac{4\pi^2}{\beta},- \frac{2\pi i\bar{\beta}}{\beta}\right)\,.
\eea

\subsection{The AdS$_3$/WCFT correspondence}
\subsubsection*{The AdS$_3$/WCFT setup--- version I}
Under CSS boundary conditions~\cite{Compere:2013bya}, asymptotically AdS$_3$ spacetimes are conjectured to be dual to WCFTs.
To be more precise, there are two different versions of the boundary conditions, demonstrated in the text and appendix of~\cite{Compere:2013bya}, respectively.
Let us first setup the holographic dictionary for WCFT in AdS$_3$ using the version in the text. In the light-like coordinate system, the BTZ black hole metric (\ref{BTZ}) with identification (\ref{id}) satisfies the CSS boundary conditions,
\begin{equation}\label{CSSbc}
g_{uv}^{(0)}=1,~~~g_{vv}^{(0)}=0,~~~\partial_vg_{uu}^{(0)}=0,~~~g_{vv}^{(2)}=T_v^2\,,
\end{equation}
with $T_v$ fixed.
The asymptotic Killing vectors obeying the boundary condition~\eqref{CSSbc} are,
\begin{equation}
\xi_n=e^{inu}\left(\partial_u-\frac{1}{2}in\partial_{\rho}\right),~~~~\eta_n=-e^{inu}\partial_v\,,
\end{equation}
The asymptotic symmetry algebra under above boundary conditions is the Virasoro-Kac-Moody algebra, which can be written as
\begin{eqnarray}\label{alg}
{}[\tilde{L}_n, \tilde{L}_m]&=&(n-m)\tilde{L}_{n+m}+\frac{c}{12}(n^3-n)\delta_{n,-m},\nonumber\\
{}[\tilde{L}_n, \tilde{P}_m]&=&-m\tilde{P}_{n+m}+m\tilde{P}_0\delta_{n,-m},\\
{}[\tilde{P}_n, \tilde{P}_m]&=&\frac{\tilde{k}}{2}n\delta_{n,-m}\,,\nonumber
\end{eqnarray}
where
\begin{equation}
c=\frac{3\ell}{2G},~~~~\tilde{k}=-\frac{\ell T_v^2}{G}\,.
\end{equation}
$\tilde{L}_0$ and $\tilde{P_0}$ are the conserved charges associate with left and right moving Killing vectors $\partial_u$ and $\partial_v$ respectively, and are related to the bulk mass and energy momentum by
$ \tilde{L}_0 ={1\over2}(E+J),~\tilde{P}_0 =-{1\over2}(E-J)$. The asymptotic charges $\tilde{ L}_n, \tilde{ P}_n$ are both finite and integrable with fixed $T_v$.
The above algebra (\ref{alg}) is not the canonical WCFT algebra (\ref{calg}). One way to relate them is through the following charge redefinition~\cite{Detournay:2012pc} \footnote{Alternatively, we can choose the field dependent asymptotic Killing vectors  following Appendix A of \cite{Apolo:2018eky},  which is a modification of the Appendix B of \cite{Compere:2013bya}. Then the bulk charges will automatically match the Virasoro-Kac-Moody generators in WCFT. In particular, for states with $\langle P_n\rangle=0,\, \forall n\neq0$, it will be obvious that $\langle L_0\rangle$ is the angular momentum.
},
\begin{equation}\label{dhhmap}
\tilde{L}_n=L_n-\frac{2P_0P_n}{k}+\frac{P_0^2\delta_{n,0}}{k},~~~~\tilde{P}_n=\frac{2P_0P_n}{k}-\frac{P_0^2\delta_{n,0}}{k}\,.
\end{equation}
In this paper, we are interested in states with $\langle P_n\rangle=0,\,  \forall n\neq0$, and this amounts to a nonlocal reparameterization of the theory,
\begin{equation}
u=x,~~~~v=\frac{ky}{2P_0}+x\,.
\end{equation}
On such states we further have the expectation values \bea\label{massspin}
 E_L &\equiv&\langle L_0\rangle=\langle\tilde{L}_0\rangle+\langle\tilde{P}_0\rangle=J,\quad {Q^2\over k}=\langle\tilde{P}_0\rangle={1\over2}(J-E)\\ \nonumber  E_L ^{inv}&\equiv&\langle L_0^{inv}\rangle=\tilde{L}_0={1\over2} (E+J),~~~~   
\eea
Note that the WCFT weight $ E_L $ corresponds to angular momentum in the bulk, while the spectral flow invariant generator $L_0^{inv}$ plays the same role of the left moving energy $L_0^{CFT}$ in a CFT$_2$. In particular, if a state has a vanishing charge, the bulk energy equals to its angular momentum which is furthermore given by the eigenvalue of $L_0$, \be E=J= E_L  , \quad if\,\, Q=0\label{zeroQ}\ee

For the BTZ black holes (\ref{BTZ}), the expectation values of the zero modes of the canonical WCFT algebra (\ref{calg}) can now be read from the background geometry,
\begin{equation}\label{DQLP}
 E_L ^{inv}=\langle L_0^{inv}\rangle=\frac{\ell}{4G} T_u^2,~~~~Q=\langle P_0\rangle=-\frac{T_v}{2}\sqrt{-\frac{\ell k}{G}}\,.
\end{equation}
The WCFT vacuum corresponds to the global AdS$_3$ with $T_{u}=T_v=-\frac{i}{2}$, and thus the vacuum has the following quantum numbers \bea\label{vacDQLP}  {E_L ^{inv}}_{vac}=-\frac{\ell}{16G} ,~~~~Q_{vac}=\frac{i}{4}\sqrt{-\frac{\ell k}{G}}\,.\eea

Here we would like to comment on a few features of the AdS$_3$/WCFT correspondence:
\begin{itemize}
\item From \eqref{DQLP}, the phase space of fixed $T_v$ is mapped to a sector with fixed charge $Q$.
On the other hand, WCFT contains  sectors with different charges, as is required by modular covariance~\cite{Detournay:2012pc}. This suggests that the phase space of WCFT consists of the union of the bulk phase spaces with different $T_v$. This interpretation was proposed by~\cite{Song:2016gtd}, and is necessary for reproducing the Bekenstein-Hawking entropy using the DHH formula. The calculations of holographic entanglement entropy~\cite{Song:2016gtd} and one-loop partition function~\cite{Castro:2017mfj} both support this interpretation.
\item It is convenient to choose a negative level $k$, which makes the charge of black holes real whereas the charge of global AdS$_3$ pure imaginary, as can be seen in eq \eqref{DQLP}. Alternatively, one could choose a positive $k$. Then global AdS$_3$ will has positive charge, and the vacuum sector will remain unitary.
\item Relatedly, a modular bootstrap with $k<0$ can be performed  assuming that all the Virasoro-Kac-Moody primary states have positive norms~\cite{Apolo:2018eky}. The negative level $k$ leads to descendent states with negative norms, whose contribution to the partition function can be estimated and is much smaller than the primaries. Furthermore, modular covariance requires that states with pure imaginary charges have to exist~\cite{Apolo:2018eky}, consistent with the fact that global AdS$_3$ has a pure imaginary charge.

\item 
One interesting question is whether the CSS boundary conditions and AdS$_3$/WCFT correspondence is consistent at the quantum level. In \cite{Castro:2017mfj}, it was shown that the one-loop determinant on BTZ background can indeed be written as a Virasoro-Kac-Moody character, as opposed to a Virasoro-Virasoro character in the AdS$_3$/CFT$_2$. Combining the result of \cite{Castro:2017mfj} and the Schottky uniformazition, it was further shown in \cite{Chen:2019xpb} that the order $c^{(0)}$ part of the R\'{e}nyi mutual information in WCFT for two interval at large distance can be obtained from a one-loop calculation in the bulk.
These results provide nontrivial checks of AdS$_3$/WCFT at the quantum level.

\end{itemize}

\subsection*{Version II}
There is an alternatively version of CSS boundary conditions as in Appendix A of \cite{Apolo:2018eky}, which is a modification of Appendix B of~\cite{Compere:2013bya}. The boundary conditions agree with (\ref{CSSbc}), with $g_{vv}^{(2)}$ still coordinate independent but allowed  to vary.
The resulting asymptotic Killing vectors are state-dependent,
\begin{equation}
\xi'_n=\xi_n+\eta_n,~~~~\eta'_n=\frac{\eta_n}{T_v}\,,
\end{equation}
from which the corresponding charges $L_n$ and $P_n$ automatically satisfy the canonical WCFT algebra (\ref{calg}) with central charge $c=3\ell/2G$ and a negative level k. Both $L_n$ and $P_n$ charges are integrable for arbitrary variations of $T_v$. These charges are related to the bulk charges \eqref{alg} in version I by the non-local map~eq \eqref{dhhmap}. Then it is straightforward to show that the zero mode of the Virasoro-Kac-Moody is just the angular momentum
\be  E_L \equiv\langle L_0\rangle= J,\quad   {Q^2\over k }={\langle P_0\rangle^2\over k}=  {1\over2}(J-E) \ee
In particular,  the BTZ black holes has the charges are $L_0=\frac{\ell}{4G}(T_u^2-T_v^2), P_0=-\frac{T_v}{2}\sqrt{-\frac{\ell k}{G}}$.
In this version of the boundary conditions, $T_v$ remains constant, but is allowed to vary. The map between the bulk and boundary is straightforward, though at the cost of field-dependent asymptotic Killing vectors. Black hole entropy, entanglement entropy, modular bootstrap can be discussed in this version, and the results  remain the same.

\section{Structure constant in AdS$_3$/WCFT}\label{sec5}
In this section, we will first derive the mean structure constant in WCFT by using the modular properties of warped conformal symmetry, and then provide a bulk calculation on the background of  BTZ black holes.
We found that the gravity picture of the mean structure constant in WCFT will still be given by Figure~\ref{plot1}. However, a crucial difference from the previous discussion is that we will use the holographic dictionary of AdS$_3$/WCFT, instead of  AdS$_3$/CFT.

\subsection{Torus one-point functions}\label{subsec5.1}
Due to the warped conformal algebra \eqref{calg}, the  operator spectrum in WCFT can be labelled by conformal weights and charges, i.e. the eigenvalues of $L_0$ and charge under $P_0$. Due to charge conservation, only chargeless operators will have non-vanishing one-point function. For a primary operator $\mathcal{O}$ with conformal weight $h_{\mathcal{O}}$ and zero charge, the one-point function on a torus \eqref{spi}~\eqref{thermalid} is defined by
\begin{equation}
\langle\mathcal{O}\rangle_{\beta, \bar{\beta}}=\mathrm{Tr}(\mathcal{O}e^{-\beta L_0+\bar{\beta}P_0})\,.
\end{equation}
Similar to the story in CFT, the warped conformal transformation (\ref{St}) which swap the spatial and thermal circle on the torus plays the role as modular transformation in WCFT. Under the warped conformal transformation the one-point function transforms as~\cite{Song:2017czq}
\begin{equation}
\langle\mathcal{O}\rangle_{\beta, \bar{\beta}}
=e^{k\frac{\bar{\beta}^2}{4\beta}}\left(\frac{\partial x'}{\partial x}\right)^{h_{\mathcal{O}}}\langle\mathcal{O}\rangle_{\frac{4\pi^2}{\beta}, -\frac{2\pi i\bar{\beta}}{\beta}}\,.\end{equation}
Here the form of the finite transformation of $\mathcal{O}$ is determined by the chiral scaling symmetry of the theory. It behaves like an $h_{\mathcal{O}}$-form under $x$-direction diffeomorphism and like a scalar under $U(1)$ transformation.

Now take limit $\beta\to0^+$, suppose the eigenvalues of $L_0$ are bounded from below, we have
\begin{equation}
\langle\mathcal{O}\rangle_{\beta, \bar{\beta}}=\langle\chi|\mathcal{O}|\chi\rangle \left(-i\frac{2\pi}{\beta}\right)^{h_{\mathcal{O}}}e^{-\frac{4\pi^2}{\beta} {E_L}_{\chi}-\frac{2\pi i\bar{\beta}}{\beta}Q_{\chi}+k\frac{\bar{\beta}^2}{4\beta}}
\end{equation}
where $|\chi\rangle$ is the lightest state with non-vanishing three-point coefficient $\langle\chi|\mathcal{O}|\chi\rangle\neq0$. The charge conservation for the three-point function requires $\mathcal{O}$ to be chargeless~\cite{Song:2017czq}. Here we use $ {E_L}_{\chi}$ for the eigenvalue of $L_0$ on the cylinder acting on $|\chi\rangle$ and $Q_{\chi}$ is the eigenvalue of $P_0$ acting on $|\chi\rangle$ which is known as the charge of $|\chi\rangle$.

On the other hand, the one-point function of the primary operator $\mathcal{O}$ can be rewritten as
\begin{equation}
\langle\mathcal{O}\rangle_{\beta, \bar{\beta}}=\int\int d E_L  dQ T_{\mathcal{O}}( E_L , Q)e^{-\beta E_L +\bar{\beta}Q}\,,
\end{equation}
where
\begin{equation}
T_{\mathcal{O}}( E_L , Q)=\sum_{i}\langle i|\mathcal{O}|i\rangle\delta(Q-Q_i)\delta( E_L - {E_L}_i)\,,
\end{equation}
is the density of states weighted by the one-point function. The integral above can be inverted to find $T_{\mathcal{O}}( E_L , Q)$,
\begin{eqnarray}
T_{\mathcal{O}}( E_L , Q)&=&\int\frac{d\beta}{2\pi}\frac{d\bar{\beta}}{2\pi}\langle\mathcal{O}\rangle_{\beta, \bar{\beta}}e^{\beta E_L -\bar{\beta}Q}\nonumber\\
&=&\int\frac{d\beta}{2\pi}\frac{d\bar{\beta}}{2\pi}\langle\chi|\mathcal{O}|\chi\rangle \left(-i\frac{2\pi}{\beta}\right)^{h_{\mathcal{O}}}e^{-\frac{4\pi^2}{\beta} {E_L}_{\chi}-\frac{2\pi i\bar{\beta}}{\beta}Q_{\chi}+k\frac{\bar{\beta}^2}{4\beta}+\beta E_L -\bar{\beta}Q}
\end{eqnarray}
At large $ E_L $ and $-Q$, this integral is dominated by a saddle point with
\begin{equation}
\beta=2\pi\sqrt{-\frac{ {{E^{inv}_L}_{\chi}}}{ E_L ^{inv}}},~~~~\bar{\beta}=\frac{4\pi}{k}\left(iQ_{\chi}+Q\sqrt{-\frac{ {E^{inv}_L}_{\chi}}{ E_L ^{inv}}}\right)\,.
\end{equation}
where $ E_L ^{inv}= E_L -\frac{Q^2}{k}$, $ {E^{inv}_L}_{\chi}= {E_L}_{\chi}-\frac{Q_{\chi}^2}{k}$. When $k$ is negative, the condition for the validity of the saddle point method is $ E_L ^{inv}\gg1$. It will become clear that negative $k$ is responsible for the $U(1)$ charge for excited states to be real when considering its gravity dual. It is easy to check that these shifted conformal dimensions $ E_L ^{inv}$ and $ {E^{inv}_L}_{\chi}$ are the eigenvalues of the zero modes of spectral flow invariant generators defined in (\ref{spfinvL}). To have the real saddle points, it is assumed that $ {E^{inv}_L }_{\chi}<0$ and $Q_{\chi}$ is pure imaginary. Under saddle point approximation, the expression of the  $T_{\mathcal{O}}( E_L , Q)$ can be written as
\begin{equation}
T_{\mathcal{O}}( E_L , Q)=\frac{(-i)^{h_{\mathcal{O}}}\langle\chi|\mathcal{O}|\chi\rangle}{\sqrt{-k {E^{inv}_L}_{\chi}}}\left(\sqrt{-\frac{ {E^{inv}_L}_{\chi}}{ E_L ^{inv}}}\right)^{2-h_{\mathcal{O}}}e^{4\pi\sqrt{- {E^{inv}_L}_{\chi} E_L ^{inv}}-\frac{4\pi i}{k}Q_{\chi}Q}\,.
\end{equation}
We can also carry out the density of states at large $ E_L $ and $-Q$ by choosing the operator $\mathcal{O}$ being the identity and setting $ {E^{inv}_L}_{\chi}\to {E^{inv}_L}_{vac}$ in $T_{\mathcal{O}}( E_L , Q)$,
\bea
\rho( E_L , Q)&=&\frac{1}{\sqrt{-k {E^{inv}_L}_{vac}}}\left(\sqrt{-\frac{ {E^{inv}_L}_{vac}}{ E_L ^{inv}}}\right)^2e^{S}\,,\\
\eea
where $S$ is the entropy formula derived in~\cite{Detournay:2012pc},
\bea
S&=&4\pi\sqrt{- {E^{inv}_L}_{vac} E_L ^{inv}}-\frac{4\pi i}{k}Q_{vac}Q\,.
\eea
As discussed in CFT, one can define the typical value of the three-point coefficient $\mathcal{C}_{\mathcal{O}}( E_L , Q)$ by
\begin{equation}
\mathcal{C}_{\mathcal{O}}( E_L , Q)\equiv\frac{T_{\mathcal{O}}( E_L , Q)}{\rho( E_L , Q)}\,.
\end{equation}
At large $ E_L ^{inv}$, $\mathcal{C}_{\mathcal{O}}( E_L , Q) $ can be approximated by \begin{eqnarray}\label{epsilon}
&&\mathcal{C}_{\mathcal{O}}( E_L , Q)\sim\\\nonumber
&&(-i)^{h_{\mathcal{O}}}\langle\chi|\mathcal{O}|\chi\rangle\sqrt{\frac{ {E^{inv}_L}_{\chi}}{ {E^{inv}_L}_{vac}}}\sqrt{-\frac{ E_L ^{inv}}{ {E^{inv}_L}_{\chi}}}^{h_{\mathcal{O}}}e^{4\pi\left(\sqrt{\frac{- {E^{inv}_L}_{\chi}}{- {E^{inv}_L}_{vac}}}-1\right)\sqrt{- {E^{inv}_L}_{vac} E_L ^{inv}}-\frac{4\pi i}{k}\left(\frac{Q_{\chi}}{Q_{vac}}-1\right)Q_{vac}Q}\,.
\end{eqnarray}
The above formula (\ref{epsilon}) characterizes the asymptotic growth of the mean three point coefficient in WCFT. This is similar to the result of~\cite{Das:2017vej} in the context of CFTs with an additional U(1) symmetry.
Although the expression (\ref{epsilon}) looks complicated, we will see that it can also be interpreted as the tadpole diagram \eqref{cbk} in the bulk gravity.

\subsection{Matching in AdS$_3$/WCFT}\label{subsec5.2}
Now we consider the bulk interpretation of the average three-point coefficient $\mathcal{C}_{\mathcal{O}}( E_L , Q)$ (\ref{epsilon}) in the context of AdS$_3$/WCFT. Under CSS boundary conditions~\cite{Compere:2013bya}, Einstein gravity on asymptotically AdS$_3$ spacetimes are dual to WCFTs.
Similar to the previous section, in this section we also consider the large $c$ limit of the heavy-heavy-light type correlators with $\mathcal{O}$ light and $| E_L ,Q\rangle$ heavy. For the light operator $\mathcal{O}$ with $1\ll h_{\mathcal{O}}\ll\frac{c}{24}$, its dual field can be viewed as perturbative bulk field lives in AdS$_3$. 
The heavy state with $ E_L ^{inv}, -\frac{Q^2}{k}\ge\frac{c}{24}$  is well described by the BTZ geometry (\ref{BTZ}) with the identification (\ref{id}).

We propose that the bulk dual of the expectation value for the torus one-point function is still given by the tadpole diagram as depicted in Figure~\ref{plot1}. Denote the bulk duals of $\mathcal{O}$ by $\phi_{\mathcal{O}}$, and $\chi$ by $\phi_\chi$. Both $\phi_{\mathcal{O}}$ and $\phi_\chi$ carry mass and spin. The amplitude contains a propagator of $\phi_{\mathcal{O}}$ emanating from infinity, a propagator of $\phi_{\chi}$ wrapping the black hole, and a cubic coupling  $\langle\chi|\mathcal{O}|\chi\rangle$. In the semiclassical limit, the propagator can be approximated by the on-shell action of massive spinning particles
\bea \label{Obk}
\mathcal{C}_{\mathcal{O}}^{bk}( E_L , Q)&=& \langle\chi|\mathcal{O}|\chi\rangle e^{-S_{\mathcal{O}}} e^{-S_\chi}\\
e^{-S_{\mathcal{O}}} &=& T_u^{\frac{\ell m_{\mathcal{O}}+s_{\mathcal{O}}}{2}}T_v^{\frac{\ell m_{\mathcal{O}}-s_{\mathcal{O}}}{2}},\quad  e^{-S_\chi}=e^{-\ell m_{\chi}2\pi(T_v+T_u)-s_{\chi}2\pi(T_u-T_v)}\nonumber
\eea
where in the second line we have used the results \eqref{bhpo} and \eqref{os}.
Note that the bulk result written in terms of the bulk variables is the same as the story of AdS$_3$/CFT$_2$ proposed in~\cite{Kraus:2016nwo} and generalized in the previous section. In the following we will use the holographic dictionary of AdS$_3$/WCFT to show that the bulk amplitude \eqref{Obk} will match the average three-point coefficient in WCFT (\ref{epsilon}) up to some normalization factors.
This indicates that holographic dualities for WCFT has another universal property, that is, the agreement between the tadpole diagram in the bulk and the mean three-point coefficient on the boundary.

\subsubsection*{The bulk dual of $\mathcal{O}$}
As discussed in the story of AdS$_3$/CFT$_2$, we expect to reproduce the contribution from the operator $\mathcal O$ by the boundary to bulk  propagator of a bulk field $\phi_{\mathcal O}$ from infinity to the horizon.
Under the WKB approximation, the propagator of a massive particle from infinity to the horizon of the BTZ metric \eqref{BTZ} is given by \eqref{bhpo}.
Now we need to use the AdS$_3$/WCFT dictionary to match the quantum numbers.

The bulk-boundary map of AdS$_3$/WCFT \eqref{massspin} relates the conformal dimension and charge to the dimensionless energy and angular momentum. For a perturbative field under the WKB approximation, it increases the dimensionless energy and angular momentum by its mass and spin, $\delta E=\ell m,\, \delta J=s.$  From \eqref{zeroQ}, we learn that the bulk dual of the chargeless operator $\mathcal{O}$ will have equal mass and spin, which is further given by its conformal weight,
\begin{equation}\label{mOWCFT}
\ell m_{\mathcal {O}}=s_{\mathcal{O}}= {E_L}_{\mathcal O}+{c\over24}=h_{\mathcal O}\,.
\end{equation}
where the shift ${c\over24}$ comes from the background.
The dual particle $\phi_{\mathcal{O}}$ is  thus a massive spinning particle with equal mass and spin. Its regularized amplitude propagating from infinity to the horizon is given by (\ref{bhpo}). 
Plugging the above relation into the general formula \eqref{bhpo} for massive spinning particles, the radial boundary to bulk propagator can be written as
\begin{equation}\label{eso}
e^{-S_{\mathcal{O}}}=T_u^{h_{\mathcal{O}}}.
\end{equation}

\subsubsection*{The bulk dual of $\chi$}
For the lightest state $|\chi\rangle$ coupled to $\mathcal{O}$, we consider the region $1\ll {E^{inv}_L}_{\chi}+\frac{c}{24}<\frac{c}{24}$, $1\ll-\frac{Q_{\chi}^2}{k}+\frac{c}{24}<\frac{c}{24}$. The bulk dual is again a massive spinning particle going around the horizon, whose on-shell action in terms of mass and spin was given in (\ref{os}). 
In the following, we use the  AdS$_3$/WCFT dictionary to further relate mass and spin to quantum numbers in WCFT.

When  $\phi_{\chi}$ is in the perturbative region, the discussion is similar to the operator $\mathcal{O}$, but with arbitrary charge $Q_\chi$ turned on. One can relate the mass and spin of $\phi_{\chi}$ to the conformal dimension and charge of $|\chi\rangle$ through (\ref{massspin}). Comparing to CFTs, $L_0^{inv}$ and $-\frac{P_0^2}{k}$ play the role as left and right moving energies, respectively. So we have the following relation,
\begin{equation}\label{bkbnp}
 {E^{inv}_L}_{\chi}+\frac{c}{24}=\frac{\ell m_{\chi}+s_{\chi}}{2},~~~~-\frac{Q_{\chi}^2}{k}+\frac{c}{24}=\frac{\ell m_{\chi}-s_{\chi}}{2}\,.
\end{equation}

In the non-perturbative region, the relation between the mass and spin in the bulk and conformal dimension and charge in WCFT can be determined through the backreacted geometry, similar to the discussion in section \ref{subsec2.3}.
For a spinning particle with mass $m_\chi$ and spin $s_\chi$, the backreacted geometry is still given by \eqref{conical},
which can be rewritten in the standard form (\ref{brg}) with the standard spatial identification  (\ref{id}), and the ``temperatures'' $T_{\chi u},\,T_{\chi v}$ determined by the mass and spin (\ref{TxuTxv}).
However, the crucial difference now is that we should use the AdS$_3$/WCFT dictionary  (\ref{massspin})  to further relate the bulk quantities $T_{\chi u},\,T_{\chi v}$ to WCFT quantum numbers. To be more explicit, we have
\be
 {E^{inv}_L }_{\chi}=\frac{\ell}{4G} T_{\chi u}^2,~~~~Q_\chi=-\frac{T_{\chi v} }{2}\sqrt{-\frac{\ell k}{G}}\
\ee
Plugging the above relation to \eqref{TxuTxv} and using (\ref{vacDQLP}), one can rewrite the mass and spin of $\phi_{\chi}$ in terms of its spectral flow invariant dimension and charge,
\begin{equation}\label{mx}
m_{\chi}=\frac{1}{4G}-\frac{1}{8G}\left(\sqrt{\frac{- {E^{inv}_L}_{\chi}}{- {E^{inv}_L}_{vac}}}+\frac{Q_{\chi}}{Q_{vac}}\right),~~~~s_{\chi}=-\frac{\ell}{8G}\left(\sqrt{\frac{- {E^{inv}_L}_{\chi}}{- {E^{inv}_L}_{vac}}}-\frac{Q_{\chi}}{Q_{vac}}\right)\,.
\end{equation}
Note that the relation in the perturbative region \eqref{bkbnp} can be obtained as the leading order expansion in the limit  $\ell m_\chi,\, s_\chi\ll c$. We will henceforth focus on the relation \eqref{mx}.  Substituting the mass and spin formula above into (\ref{os}), we find the contribution to amplitude from $\phi_{\chi}$ can be recast as
\begin{equation}\label{esx}
e^{-S_{\chi}}=e^{\frac{\pi\ell}{2G}\left(\sqrt{\frac{- {E^{inv}_L}_{\chi}}{- {E^{inv}_L}_{vac}}}-1\right)T_u+\frac{\pi\ell}{2G}\left(\frac{Q_{\chi}}{Q_{vac}}-1\right)T_v}\,.
\end{equation}

\subsubsection*{The bulk dual for the structure constant}
Putting the two parts (\ref{eso}) and (\ref{esx}) together, we can rewrite the total amplitude \eqref{Obk} for the process given by Figure~\ref{plot1} under CSS boundary conditions,
\begin{equation}\label{twoparts}
\mathcal{C}^{bk}_{\mathcal{O}}( E_L , Q)=\underbrace{\langle\chi|\mathcal{O}|\chi\rangle}_{\text{vertex}}\, \underbrace{T_u^{h_{\mathcal{O}}}}_{\langle\phi_{\mathcal{O}}\phi_{\mathcal{O}}\rangle}\,
\underbrace{e^{\frac{\pi\ell}{2G}\left(\sqrt{\frac{- {E^{inv}_L}_{\chi}}{- {E^{inv}_L}_{vac}}}-1\right)T_u+\frac{\pi\ell}{2G}\left(\frac{Q_{\chi}}{Q_{vac}}-1\right)T_v}}_
{\langle\phi_{\chi}^{\dagger}\phi_{\chi}\rangle}\,,
\end{equation}
Up to an overall normalization that does not depend on the $ E_L $ and $Q$, this amplitude recover the WCFT result (\ref{epsilon}) rewritten in terms of temperatures.

For both AdS$_3$/CFT$_2$ and AdS$_3$/WCFT,  the bulk dual of the asymptotic growth of three point coefficient can be written as a tadpole diagram~\eqref{cbk}, consisting of the vertex, the propagator $\langle\phi_{\mathcal{O}}\phi_{\mathcal{O}}\rangle$ from boundary to the horizon, and the propagator $\langle\phi_\chi\phi_{\chi}\rangle$ around the horizon. However, the general formula \eqref{cbk} needs to take different parameters in the two cases.
In WCFT, all states in the highest weight representation of Virasora-Kac-Moody algebra are eigenstates of the $U(1)$ generator $P_0$. Charge eigenstate together with charge conservation impose strong constraints in the bulk~\cite{Song:2017czq}. For instance, charge conservation requires that  the operator $\mathcal{O}$ to be chargeless. In the semiclassical bulk picture, this means that the massive particle has equal mass and spin \eqref{mOWCFT}, which were not discussed in the original AdS$_3$/CFT$_2$ calculation for scalar operators.  In the generalized version, namely section \ref{sec3} of this paper, particles with equal mass and spin in AdS$_3$ map to operators with $\bar h=0$ in CFT$_2$, which is not in the semiclassical region $1\ll h_{\mathcal O}, \bar h_{\mathcal O}\ll {c\over24}$.
 More explicitly, there is no $T_v$ dependence for  $\langle\phi_{\mathcal{O}}\phi_{\mathcal{O}}\rangle$ in WCFT, as opposed to the CFT result  \eqref{bkAdSCFT} which is only valid for non-vanishing $\bar h_{\mathcal{O}}$.
Another interesting feature in the AdS$_3$/WCFT formula \eqref{twoparts} is in the contribution from $\phi_{\chi}$.
The $|\chi\rangle$ state in the dual WCFT is a state with pure imaginary charge which is predicted by the modular bootstrap of WCFT~\cite{Apolo:2018eky}. $Q_{\chi}$ and $Q_{vac}$ are both pure imaginary numbers and the amplitude is real.

To summarize, we derived a universal formula for the asymptotic growth of the average three-point coefficient, and performed a bulk calculation in the context of AdS/WCFT correspondence in the semiclassical region.
In addition to the fact that the thermal entropy of a BTZ black hole is captured by the DHH formula in WCFT~\cite{Detournay:2012pc}, the matching of (\ref{cbk}) and (\ref{epsilon}) gives further evidence of the black hole geometries in asymptotically AdS$_3$ spacetimes can emerge upon coarse-graining over microstates in WCFTs.
In this paper, we only considered the semiclassical region and the contribution from the primary operators. It is interesting to further perform a geodesic Witten type calculation in AdS/WCFT, parallel to the calculation of \cite{Kraus:2017ezw} in AdS$_3$/CFT. Such a calculation will further check the consistency of CSS boundary conditions and the proposed AdS$_3$/WCFT duality at the quantum level apart from \cite{Chen:2019xpb, Castro:2017mfj}.

\section*{Acknowledgement}
We are grateful to Luis Apolo, Pankaj Chaturvedi, Bin Chen, St\'{e}phane Detournay, Pengxiang Hao, Junjie Zheng for helpful discussions.
We especially thank Alex Maloney for bringing the question to us during the Aspen workshop. This work was partially supported by the National Thousand-Young-Talents Program of China and NFSC Grant No. 11735001, and the Fundamental Research Funds for the Central Universities No. 2242019R10018. The authors thank the Tsinghua Sanya International Mathematics Forum for hospitality during the workshop and research-in-team program ``Black holes, Quantum Chaos, and Solvable Quantum System'' and ``Black holes and holography''.  W.S. would also like to thank the workshop ``Quantum Gravity and New Moonshines'' and ``Information in Quantum Field Theory'' at the Aspen Center for Physics, which is supported by Simons Foundation and National Science Foundation grant PHY-1066293.

\section*{Appendix A: an alternative derivation of the on-shell action at the horizon}
In this appendix, we will use the non-rotating frame to the calculate the on-shell action for particle $\phi_{\chi}$.

Since the horizon is a degenerate hypersurface, here we use the trick to calculate the on-shell action. We take the orbit of the particle $\phi_{\chi}$ to be a circular orbit with constant radius greater than the horizon's. It can be shown that the constant radius orbits outside the horizon are no longer geodesics, but with accelerations. Then we evaluate the action of $\phi_{\chi}$ on that orbit, this is not on-shell since (\ref{MPD0}) is not satisfied. Finally, we take the radius of that orbit tends to horizon. On the horizon, the worldline action become on-shell, and we will get the desired result.

By using the metric (\ref{BTZ}), we can find out the tangent and two normal vectors of the constant radius obit outside the horizon at constant time,
\begin{eqnarray}
\dot{X}^{\mu}\partial_{\mu}&=&\frac{1}{\ell\sqrt{2\rho+T_u^2+T_v^2}}\partial_u+\frac{1}{\ell\sqrt{2\rho+T_u^2+T_v^2}}\partial_v,\label{Xx}\\ \tilde{n}^{\mu}\partial_{\mu}&=&\frac{2\sqrt{\rho^2-T_u^2T_v^2}}{\ell}\partial_{\rho}\,,\label{ntx}\\
n^{\mu}\partial_{\mu}&=&\frac{\rho+T_v^2}{\ell\sqrt{(\rho^2-T_u^2T_v^2)(2\rho+T_u^2+T_v^2)}}\partial_u-\frac{\rho+T_u^2}{\ell\sqrt{(\rho^2-T_u^2T_v^2)(2\rho+T_u^2+T_v^2)}}\partial_v\,.\nonumber\\\label{nx}
\end{eqnarray}
Similar to the discussion of particle $\phi_{\mathcal{O}}$, the spin contribution can be written as
\begin{equation}\label{Sanom1}
S_{\chi}^{\mathrm{spin}}=s_{\chi}\log\left(\frac{\tilde{q}(\tau_f)\cdot\tilde{n}_f-q(\tau_f)\cdot\tilde{n}_f}{\tilde{q}(\tau_i)\cdot\tilde{n}_i-q(\tau_i)\cdot\tilde{n}_i}\right)\,,
\end{equation}
Before taking the limit, the orbit has acceleration, the notion of parallel transport should be replaced by Fermi-Walker transport. Here the vectors $q^{\mu}$ and $\tilde{q}^{\mu}$ satisfying
\begin{eqnarray}\label{FW}
&&q^2=-1,\quad \tilde{q}^2=1,\quad q\cdot\tilde{q}=0,\quad q\cdot\dot{X}=\tilde{q}\cdot\dot{X}=0,\nonumber\\
&&\nabla q^{\mu}=-\dot{X}^{\mu}(q\cdot A),\quad \nabla\tilde{q}^{\mu}=-\dot{X}^{\mu}(\tilde{q}\cdot A)\,,
\end{eqnarray}
where $A^{\mu}=\nabla\dot{X}$ is the acceleration corresponds to the tangent vector $\dot{X}$. The Fermi-Walker transport was imposed to ensure that the frame $(q(\tau), \tilde{q}(\tau))$ is non-rotated along the orbit, and compare to another frame $(n(\tau), \tilde{n}(\tau))$, we can obtain the relative change of the Lorentz boost. The explicit expressions of $q$ and $\tilde{q}$ for constant radius orbit outside the horizon can be written down by solving (\ref{FW}),
\begin{eqnarray}
q^{\mu}\partial_{\mu}&=&\frac{(\rho+T_v^2)\cosh\left(\frac{(T_u^2-T_v^2)\varphi}{\sqrt{2\rho+T_u^2+T_v^2}}\right)}{\ell\sqrt{(\rho^2-T_u^2T_v^2)(2\rho+T_u^2+T_v^2)}}\partial_u-\frac{(\rho+T_u^2)\cosh\left(\frac{(T_u^2-T_v^2)\varphi}{\sqrt{2\rho+T_u^2+T_v^2}}\right)}{\ell\sqrt{(\rho^2-T_u^2T_v^2)(2\rho+T_u^2+T_v^2)}}\partial_v\nonumber\\
&&-\frac{2\sqrt{\rho^2-T_u^2T_v^2}\sinh\left(\frac{(T_u^2-T_v^2)\varphi}{\sqrt{2\rho+T_u^2+T_v^2}}\right)}{\ell}\partial_{\rho}\,,\label{qx}\\
\tilde{q}^{\mu}\partial_{\mu}&=&-\frac{(\rho+T_v^2)\sinh\left(\frac{(T_u^2-T_v^2)\varphi}{\sqrt{2\rho+T_u^2+T_v^2}}\right)}{\ell\sqrt{(\rho^2-T_u^2T_v^2)(2\rho+T_u^2+T_v^2)}}\partial_u+\frac{(\rho+T_u^2)\sinh\left(\frac{(T_u^2-T_v^2)\varphi}{\sqrt{2\rho+T_u^2+T_v^2}}\right)}{\ell\sqrt{(\rho^2-T_u^2T_v^2)(2\rho+T_u^2+T_v^2)}}\partial_v\nonumber\\
&&+\frac{2\sqrt{\rho^2-T_u^2T_v^2}\cosh\left(\frac{(T_u^2-T_v^2)\varphi}{\sqrt{2\rho+T_u^2+T_v^2}}\right)}{\ell}\partial_{\rho}\,.\label{qtx}
\end{eqnarray}
The metric (\ref{BTZ}) has identification as (\ref{id}), so let us set the initial point with $\varphi=0$ and final point with $\varphi=2\pi$, the difference in $q$ and $\tilde{q}$ between initial and final point are the spin effects. Substituting the formulas (\ref{qx}), (\ref{qtx}), and (\ref{ntx}) into (\ref{Sanom1}), we can get,
\begin{equation}
S_{\mathrm{spin}}=s_{\chi}\frac{2\pi(T_u^2-T_v^2)}{\sqrt{2\rho+T_u^2+T_v^2}}\,.
\end{equation}
Now we can take limit $\rho\to T_uT_v$ to get the spin contribution on-shelly and combine with the first part of (\ref{worldlineacchicft}), we can write down the on-shell action of particle $\phi_{\chi}$.
\begin{equation}
S_{\chi}=\ell m_{\chi}2\pi(T_v+T_u)+s_{\chi}2\pi(T_u-T_v)\,.
\end{equation}

\section*{Appendix B: conical defects with spin}
In this appendix, we will check that \eqref{conical} is indeed a solution in three dimensional spacetime with negative cosmological constant, sourced by  a point particle with mass $m$ and spin $s$ at the origin,
\bea \label{eeq}\mathcal{G}_{\mu\nu}&=&8\pi G T_{\mu\nu}\,, \\
T^{00}&=&m\delta^2(x),~~~~x^iT^{0j}-x^jT^{0i}=s\epsilon^{ij}\delta^2(x)\,,\label{source}
\eea
where $\mathcal{G}_{\mu\nu}\equiv R_{\mu\nu}-{1\over2}Rg_{\mu\nu}-{1\over\ell^2} g_{\mu\nu}$.
The indices $(0, i, j)$ labels the local Cartesian coordinates $(x^0,x^1,x^2)$.
One can directly check that \eqref{conical} indeed satisfies the above Einstein equation \eqref{eeq} following the lines of arguments of \cite{Deser:1983tn,Deser:1983nh}.
Alternatively, by integrating both sides of \eqref{eeq} over a Cauchy surface,  \eqref{source} indicates that the mass and spin can be calculated from the quasi-local gravitational charges. In the covariant formalism, charges associated to exact Killing vectors can be calculated anywhere, and  the difference between the usual ADM charges and quasi-local charges comes from the choice of Killing vectors~\cite{Iyer:1994ys}. In general charge integrability imposes strong constraints on the normalizations.
More explicitly,  the parameters of the local source can be calculated from the covariant charges \be  m= Q[\p_0],\quad s C=Q[\p_\theta], \ee
where  $x^1=r \cos C\theta, \,x^2=r \sin  C\theta, \, x^0=t-A \theta$ with $\theta\sim\theta +2\pi$ are the locally polar coordinates. Explicit expression of the charges can be found in~\cite{Iyer:1994ys}.

Now let us check the local source of the conical solution \eqref{conical} using the prescription above.
Consider the a conical defect solution in global AdS with ``jump''
\bea\label{ds2rphi}
{ds^2\over\ell^2}&=&{dr^2\over 1+r^2}+r^2d\varphi^2-(1+r^2)(dt-{A\over C} d\varphi)^2\,.\\
\varphi&\sim&\varphi+2\pi C,\quad  C=1 -{\delta\varphi\over2\pi},
\eea
which can be brought to the standard form \eqref{brg} where we can read the asymptotic charges by, \bea u={\varphi\over C}+{t\over C-A},\quad v={\varphi\over C}-{t\over C+A}\,.\eea
The total gravitational energy and angular momentum are measured using the asymptotic Killing vectors $ E_L  =Q[\p_u]=-{\ell\over16G}(C-A)^2,\quad  E_R =Q[-\p_v] =-{\ell\over16G}(C+A)^2$.
On the other hand, the local source can be measured by the Killing vectors normalized properly near the origin, using
the locally Cartesian coordinates $x^0=\ell (t-{A\over C}\varphi),\, x^1=r\cos\varphi, \, x^2= r\sin\varphi$.
Therefore the particle mass can be calculated by considering the infinitesimal charge
\bea
\ell \delta m&=& \delta Q[\p_t]\nonumber\\
&=& {1\over C-A}\delta Q[\p_u]+  {1\over C+A}\delta Q[\p_v]\nonumber\\
&=&-\ell {\delta C\over4G}\,.
\eea
By requiring that smoothness when $m=0$, we can integrate the above charge and get the particle mass
\be m={1-C\over 4G}={\delta\varphi \over 8\pi G}\,.\ee
Similarly,
we can get the spin of the particle by
\bea s &=& {1\over C} Q[C\p_\varphi ]={1\over C} (Q[\p_u]+Q[\p_v])={ A\ell \over 4G}\,.
\eea

\end{document}